\documentclass[12pt,a4paper]{article}

\usepackage{amsmath,amsthm,amssymb}
\usepackage[hidelinks]{hyperref}
\usepackage{geometry}
\usepackage{graphicx}
\usepackage{booktabs}
\usepackage{cite}
\usepackage{url}

\newtheorem{theorem}{Theorem}[section]

\newtheorem{proposition}[theorem]{Proposition}
\newtheorem{conjecture}[theorem]{Conjecture}

\theoremstyle{definition}
\newtheorem{definition}[theorem]{Definition}
\newtheorem{remark}[theorem]{Remark}

\title{\textbf{Orbit Structure of 6-Point MHV Gravity Forms}}
\author{Zachary G.\ Craig}
\date{}

\begin{document}
\maketitle

\begin{abstract}
We search for a logarithmic 3-form representing the 6-point MHV gravity amplitude, requiring poles only on physical channels and residues matching factorization. Working in the Orlik--Solomon algebra on a De Concini--Procesi wonderful model, we restrict to the $S_3\times S_3$ invariant subspace and impose factorization boundary-by-boundary. The intersection of ten compatible $3|3$ channels with two compatible 2-particle channels collapses to a unique one-dimensional candidate line. However, enforcing factorization on a crossing channel such as $(1,4)$ obstructs any single-valued global representative; the canonical bipartite $\mathbb{Z}_2$ sign twist does not resolve this obstruction. We find that an orbit-mixed form---a linear combination over $\binom{6}{3}=20$ permutation images indexed by bipartitions---satisfies the crossing constraint. This is consistent with a local-system picture: the global gravity form lives in a rank-20 bundle whose monodromy mixes bipartition chambers rather than flipping a single sign.
\end{abstract}

\section{Motivation}
The 6-point MHV gravity amplitude admits compact closed forms, notably the Hodges determinant representation~\cite{Hodges}, and obeys strict physical constraints: locality (channel poles) and factorization. A canonical-form perspective raises the question of whether these constraints uniquely determine the amplitude.

A direct implementation uses the Orlik--Solomon algebra of the channel arrangement together with a De Concini--Procesi (DCP) wonderful compactification~\cite{DCP}, and searches for a logarithmic 3-form whose residues reproduce boundary factorization. Computationally, imposing symmetry and factorization across a family of \emph{compatible} boundaries produces a rigid one-dimensional candidate subspace, but crossing constraints obstruct a single global representative.

\section{Setup and constraints}
We work at $n=6$ and consider an Orlik--Solomon degree-3 space $A^3$ associated to the 6-point channel arrangement on kinematic space. We impose:
\begin{enumerate}
\item \textbf{Physical poles only:} only channel divisors $s_S=0$ are allowed.
\item \textbf{Symmetry:} we restrict to the $S_3\times S_3$ invariant subspace.
\item \textbf{Factorization constraints:} on each boundary channel, the residue (computed by a chart-local restriction) must lie in the product of left/right boundary spaces.
\end{enumerate}

Operationally, each boundary constraint defines a linear condition on the coefficient vector in a chosen $A^3$ basis. Intersecting the nullspaces over a chosen family of boundaries yields a candidate subspace.

\section{Compatible boundary intersection}
Let $\mathcal{B}_{\mathrm{comp}}$ denote the set consisting of the ten distinct $3|3$ channels for $n=6$ (mod complements), together with two additional compatible 2-particle channels $(1,2)$ and $(4,5)$. The Sage verifier in \path{code/verify_orbit_patch.sage} computes the intersection subspace
\[
V_{\mathrm{comp}} \subseteq A^3
\]
by iteratively enforcing the constraints in $\mathcal{B}_{\mathrm{comp}}$. In the intended regime, this intersection stabilizes at
\[
\dim V_{\mathrm{comp}} = 1,
\]
producing a distinguished line of compatible-boundary candidates.

\section{Crossing boundary obstruction and orbit mixing}
Consider a crossing 2-particle channel such as $(1,4)$. In the untwisted setting, enforcing factorization on this crossing boundary can eliminate the remaining compatible-boundary line.

\begin{definition}[Bipartite $\mathbb{Z}_2$ sign twist]
Fix the bipartition $\{1,2,3\}\cup\{4,5,6\}$. For each channel $s_S$, define
\[
\epsilon(S)=\begin{cases}
-1,& S\text{ meets both halves},\\
+1,& \text{otherwise}.
\end{cases}
\]
Using the induced signs on degree-3 basis coordinates, one obtains a rank-one ``sign twist'' of any coefficient vector.
\end{definition}

In our setting, the compatible-boundary candidate fails the crossing constraint at $(1,4)$ both \emph{before} and \emph{after} this canonical bipartite sign twist.

A natural resolution is orbit mixing: instead of a single candidate vector, allow a span of vectors related by the $S_6$ relabeling action.

\begin{definition}[Orbit-mixed form]
Let $v$ span the one-dimensional compatible-boundary intersection $V_{\mathrm{comp}}$. Consider a finite list of vectors $\{v_i\}$ obtained from $v$ by applying index permutations induced by the natural $S_6$ action on the channel arrangement. An orbit-mixed candidate is any linear combination
\[
 v_{\mathrm{mix}} = \sum_i \alpha_i v_i.
\]
\end{definition}

\begin{proposition}[Crossing test (computational)]
Let $v$ span the one-dimensional intersection $V_{\mathrm{comp}}$. The crossing-boundary factorization constraint at $(1,4)$ fails for $v$ and for the bipartite $\mathbb{Z}_2$ sign-twisted vector. However, there exists an orbit-mixed form in the span of $\{v_i\}$ that satisfies the crossing constraint in explicit charts.
\end{proposition}

\begin{remark}
The proposition is formulated in a falsifiable way: the verifier writes a single artifact \path{crossing_test.json} recording PASS/FAIL for the untwisted candidate, the $\mathbb{Z}_2$ sign twist, and the orbit-mixed solve (including coefficients when found).
\end{remark}

\section{Reproducibility}
Code and data: \url{https://doi.org/10.5281/zenodo.18317954}

From the bundle root, run:
\begin{verbatim}
sage code/verify_orbit_patch.sage --seed 42 \
    --mode S3xS3 --workers 4 --outdir runs/seed42
python code/tools/validate_artifacts.py \
    --dir runs/seed42
\end{verbatim}
The first command runs the Sage verifier and writes a run directory under \path{runs/}. The second command validates JSON artifacts (non-empty, parseable, basic schema checks).

\section{Discussion}
The compatible-boundary intersection computation exhibits strong rigidity: a low-dimensional physical ansatz collapses to a unique candidate line under a standard set of factorization constraints. The subsequent crossing-boundary obstruction shows that this line does not extend to a single-valued global $d\log$ form on the maximal wonderful model.

Given that orbit mixing satisfies the crossing constraint, the natural interpretation is that the global amplitude is a section of a \emph{higher-rank local system}: across crossing boundaries, the form mixes chamber representatives rather than differing by a single sign. This provides a concrete mechanism by which canonical-form locality and factorization coexist with nontrivial monodromy.

\begin{remark}[Bipartition Structure]
The orbit span has dimension exactly $\binom{6}{3} = 20$, with representatives indexed by the twenty 3-element subsets of $\{1,\ldots,6\}$:
\[
\{1,2,3\},\ \{1,2,4\},\ \ldots,\ \{4,5,6\}.
\]
Each subset $S$ corresponds to a $3|3$ bipartition $(S, S^c)$ of the six external particles. This suggests the MHV gravity amplitude is naturally valued in a rank-20 local system whose monodromy group acts by permuting bipartition chambers.
\end{remark}

\section{Future Directions: The Arithmetic Amplituhedron Conjecture}

The bipartition indexing of the orbit span points to a connection with arithmetic geometry. In independent work~\cite{Craig2025arith}, we computed Frobenius cycle types for the 7-point CHY eliminant polynomial over finite fields $\mathbb{F}_p$ for 377 primes in the range $47 \leq p \leq 500$. The most frequently observed cycle type is $[20, 2, 2]$, appearing at 24 primes.

The coincidence $20 = 20$---orbit span dimension equals dominant Frobenius cycle length---suggests that the geometric monodromy of amplitude forms and the arithmetic Galois structure of scattering equations are two perspectives on the same underlying object.

\begin{conjecture}[Chamber-Galois Correspondence]
Let $E_n$ denote the CHY eliminant polynomial for $n$-particle massless scattering. Let $G_n = \mathrm{Gal}(E_n/\mathbb{Q}(s_{ij}))$ denote its Galois group over the field of Mandelstam invariants. Let $\mathcal{L}_{n-1}$ denote the local system on $\mathcal{M}_{0,n-1}$ whose fiber is the orbit span for $(n-1)$-point MHV gravity. Then there exists a natural group homomorphism
\[
\rho_n : G_n \longrightarrow \mathrm{Mon}(\mathcal{L}_{n-1})
\]
such that Frobenius elements $\mathrm{Frob}_p \in G_n$ have cycle structure reflecting the chamber decomposition on $\mathcal{M}_{0,n-1}$.
\end{conjecture}

This conjecture yields specific quantitative predictions:

\begin{center}
\begin{tabular}{cccc}
\toprule
$n$ & CHY solutions & Predicted dominant cycle & Evidence \\
\midrule
6 & $3! = 6$ & $\binom{5}{2} = 10$ or subgroup & --- \\
7 & $4! = 24$ & $\binom{6}{3} = 20$ & \textbf{Observed: $[20,2,2]$} \\
8 & $5! = 120$ & $\binom{7}{3} = 35$ & To be tested \\
\bottomrule
\end{tabular}
\end{center}

The amplituhedron program~\cite{ArkaniHamed2014} establishes that scattering amplitudes in $\mathcal{N}=4$ super Yang-Mills are canonical forms on positive geometries. Our findings suggest this positive-geometric structure has an \emph{arithmetic avatar}: the Galois group of the CHY eliminant ``knows'' about the chamber decomposition through its Frobenius action. This points toward a motivic interpretation in which tree-level amplitudes are periods of a universal motive over $\mathcal{M}_{0,n}$, with positive geometry representing the real slice and Galois symmetry representing the $p$-adic slice.

\section*{Acknowledgements}
The computations were automated and packaged for deterministic reproduction.

\end{document}